\documentclass{MIPRO}

\usepackage{cite}
\usepackage{amsmath,amssymb,amsfonts}
\usepackage{algorithmic}
\usepackage{graphicx}
\usepackage{textcomp}
\usepackage{xcolor}
\usepackage[T1]{fontenc}
\usepackage{flushend}
\usepackage{multirow}
\usepackage{multicol}
\usepackage{array}
\usepackage{url}
\usepackage{amsmath}
\usepackage{bm}
\usepackage{siunitx}
\usepackage{hyperref}
\usepackage[capitalise,nameinlink]{cleveref}
\usepackage{glossaries}

\newacronym{pde}{PDE}{partial differential equation}

\newcommand{\CC}{C\nolinebreak\hspace{-.05em}\raisebox{.2ex}{\small\bf +}\nolinebreak\raisebox{.2ex}{\small\bf +}}

\begin{document}

\title{Load Balanced Parallel Node Generation for Meshless Numerical Methods}

\author{
\IEEEauthorblockN{
J. Vehovar\IEEEauthorrefmark{1}\IEEEauthorrefmark{2},
M. Rot\IEEEauthorrefmark{1}\IEEEauthorrefmark{2},
M. Depolli\IEEEauthorrefmark{2},
G. Kosec\IEEEauthorrefmark{2}
}

\IEEEauthorblockA{\IEEEauthorrefmark{1} 
Jozef Stefan International Postgraduate School, Jamova cesta 39, 1000 Ljubljana, Slovenia}

\IEEEauthorblockA{\IEEEauthorrefmark{2} 
Institut "Jožef Stefan", Parallel and Distributed Systems Laboratory, Jamova cesta 39, 1000 Ljubljana, Slovenia}

jon.vehovar@ijs.si

}

\maketitle

\begin{abstract}
Meshless methods are used to solve partial differential equations by approximating differential operators at a node as a weighted sum of values at its neighbours. One of the algorithms for generating nodes suitable for meshless numerical analysis is an n-dimensional Poisson disc sampling based method. It can handle complex geometries and supports variable node density, a crucial feature for adaptive analysis. We modify this method for parallel execution using coupled spatial indexing and work distribution hypertrees. The latter is prebuilt according to the node density function, ensuring that each leaf represents a balanced work unit. Threads advance separate fronts and claim work hypertree leaves as needed while avoiding leaves neighbouring those claimed by other threads. Node placement constraints and the partially prebuilt spatial hypertree are combined to eliminate the need to lock the tree while it is being modified. Thread collision handling is managed by the work hypertree at the leaf level, drastically reducing the number of required mutex acquisitions for point insertion collision checks. We explore the behaviour of the proposed algorithm and compare the performance with existing attempts at parallelisation and consider the requirements for adapting the developed algorithm to distributed systems.
\end{abstract}

\renewcommand\IEEEkeywordsname{Keywords}

\section{Introduction}

Meshless methods \cite{Liu2002} are a class of numerical methods used to solve \glspl{pde} when they are not solvable in closed form. Their advantage over established methods lies in their ability to solve problems on arbitrarily shaped domains without requiring any information about the connectivity between nodes that discretise the domain on which the solution is being computed on. The approach also allows for node density variability, so that areas of the domain where a more accurate solution is needed or greater solution stability is required can be discretised more finely \cite{Oanh2017}.

While it is possible for these methods to produce solutions on a completely randomly generated set of nodes, the solution quality improves if the node set satisfies certain constraints such as a lower limit on the minimal distance between nodes and local regularity \cite{Slak2019}.

There are multiple classes of methods that produce nodes suitable for use with meshless methods, such as iterative relaxation, mesh-based, sphere-packing, or advancing front algorithms \cite{Fornberg2015}. An example of the latter is the algorithm presented in \cite{Slak2019} where the authors start with seed points around which they generate new points to repeat the same process with. Their algorithm is, while performant, sequential and therefore of limited use for applications that demand rediscretisation, such as adaptive methods, which iteratively determine node density requirements for a specific problem based on error indicators \cite{Bacer2025}. In this case, a parallel version is advantageous in terms of execution time.

For clarity, we choose to hereinafter refer to objects that are in the meshless community known as nodes, where the numerical solution is computed in, as points to distinguish them from elements of tree data structures, to which we will refer to as nodes.

The previously mentioned algorithm was adapted for parallel execution and presented in \cite{Depolli2022} (in this work, we refer to it as Pfill) where the authors use a two-level spatial indexing structure. The top level consists of cells defined by seed points, which were generated using the sequential algorithm on a coarser level. A point belongs to the cell to whose generating point it is closest to. Each cell contains a $k$-d tree that further indexes the points. When multiple threads insert points into this spatial index, they use a readers-writer lock for the subtree into which they are inserting to permit multiple parallel readers of the subtree data while reserving the right to alter its data for themselves.

In this paper, we propose a new parallel advancing front algorithm that uses a $d$-dimensional hypertree (to which we hereinafter refer to as ``tree'' for brevity) for spatial indexing. The main benefit of the new approach being a lightweight and more fine-grained locking mechanism to parallelise the algorithm from \cite{Slak2019} with the aim of improving upon the performance of Pfill. We run performance tests for both algorithms in a simple disc domain and present the analysis of the results.

We present the sequential algorithm being parallelised, the parallelisation approach, its implementation, and experimental setup in \cref{sec:methods}. Experimental results and their analysis are presented in \cref{sec:results}. They are further discussed in \cref{sec:discussion}, and, finally, we state our conclusions in \cref{sec:conclusion}.

\section{Methods}
\label{sec:methods}

\subsection{Sequential algorithm}

The sequential algorithm we are adapting for parallelisation is a modified version of Poisson disc sampling described in \cite{Slak2019}. This method yields a quasi-uniform distribution of points within the domain $\Omega$ that is being filled, usually by starting from a set of seed points at its border. The algorithm also requires a spacing function $h : \mathbb{R}^d \to (0, \infty)$ that determines the spacing between points within $\Omega$ and a characteristic function $\chi_\Omega: \mathbb{R}^d \to \{0,1\}$ that is used to determine whether a point $\bm{p}$ lies within $\Omega$. The seed points represent a set of starting expansion points $\{\bm{p_i}\}_i$ that form the expansion front. The algorithm fills the domain by removing a point $\bm{p}_i$ from the expansion set and generating candidate points around it at a distance $h(\bm{p}_i)$. Next, it checks these candidates one after another to see whether they lie within $\Omega$ and whether any point already inserted into the domain is within a distance of $h(\bm{p}_i)$. If there is no such point, the candidate is inserted into the domain and added to the expansion set. The algorithm repeats this iteration until the expansion set is exhausted.

\subsection{Parallelisation concept}
\label{sec:parallelisation_concept}

Our approach is based on the tree data structure described in \cite{Rojc2022} and the supporting virtual memory data structure described in \cite{Rojc2021}. In the former, they already parallelise the previously mentioned point placement algorithm, but they use locking on the level of leaves, requiring one exclusive mutex acquisition and potentially multiple shared mutex acquisitions per point insertion. We adopt the main components of their data structure, i.e. the virtual memory vectors for both tree node types (internal and leaves) and elements, but otherwise diverge from their implementation.

We remove thread synchronisation mechanisms from the spatial tree and instead construct a work tree that is a subtree\footnote{Departing from the common usage of the term, this subtree doesn't share leaves with the parent tree and terminates earlier than the root node, but starts at the root node and terminates before reaching the parent's leaves.} of the spatial indexing tree, to which we hereinafter refer to as work tree and spatial tree, respectively, for brevity. The work tree is constructed in the initialisation phase. It's depth is recursively determined by estimating the number of points within the leaf using $h$, splitting it the estimated number of inserted points is above the work tree leaf size threshold.

The role of work tree leaf cells (hereinafter work cells) is to prevent race conditions in the spatial tree while the domain is being filled. This is ensured by three conditions. Firstly, if a work cell contains points (i.e. they are within its bounding box while the actual point coordinates are stored in the spatial tree) in the expansion queue of one thread, there may be no such points owned by other threads in any of the cell's nearest neighbours. Here, a nearest neighbour cell is any cell that shares any vertex with the central cell (we use the algorithm from \cite{Ibaroudene1991} for efficient computation of all nearest neighbours). Secondly, threads may place points outside the work cell that contains the point being expanded, but those points may not be inserted onto the expansion queue if the cell that contains the newly inserted point doesn't satisfy the first condition\footnote{Strictly speaking, some additional constraints are needed to ensure that these points don't violate the proximity condition, but the cases where this happens are rare in practice. If they occur, they can be fixed trivially.}. Lastly, all nearest neighbours of a work cell within the spatial tree must be at least one level lower than it.

Combined, these conditions ensure that any two threads have at least one work cell of separation, and that the spatial tree is subdivided enough that threads may also insert points into leaves contained within these buffer cells, but that no two threads may insert them into the same leaf at the same time.

To ensure that the algorithm indeed fills the whole domain like the sequential algorithm\footnote{The sequential algorithm requires $h$ to be smaller than the characteristic size of the domain's features being sampled.}, there needs to be a (global) set of work cells to be revisited when threads exhaust their fronts called the \emph{restart set}. Cells are inserted into it when the second condition is enforced. Since it is best for threads to revisit cells in the order they were inserted to maximise the chance of a successful claim, this set should be ordered and used as a FIFO queue. Cells are removed from this set either if a thread successfully advances its front into a cell, or restarts filling the domain from it. Before restarting from a cell, all points within it are inserted onto a thread's expansion queue, leading to a natural resumption of a front that was terminated. The termination condition for our algorithm is therefore that all threads' expansion queues must be exhausted and that there may be no cells in the restart set.

With all these considerations, three fundamental cell states can be defined. The first two conditions induce an \emph{initial state} which encompasses those cells that don't contain points of any thread's expansion queue and a \emph{claimed} (by a certain thread) state which encompasses those that do contain such points. Cells in the restart set are in an \emph{enqueued} state.

\subsection{Complexity}

The complexity of the sequential algorithm \cite{Slak2019} is $\mathcal{O}(N\log N)$ with respect to the number of inserted points $N$. This is determined by the need to traverse a spatial index tree with the depth $\mathcal{O}(\log N)$ for each candidate check where the number of checked candidates is $\mathcal{O}(N)$. As our algorithm doesn't alter this step, this is the best possible computational complexity we can achieve. Placement of seed points is independent of the problem size, leaving the prebuilding step as a possibility of increased complexity. The number of tree nodes to be created is $\mathcal{O}(N)$ and the work cell neighbour search requires traversal of $\mathcal{O}(\log N)$ tree levels for $\mathcal{O}(N)$ work cells, yielding $\mathcal{O}(N\log N)$ complexity. Thus, our algorithm's complexity remains $\mathcal{O}(N\log N)$.

\subsection{Implementation}

We choose to implement the algorithm that is about to be described in \CC{} where we use the \CC{} standard library's \texttt{std::thread} to generate multiple parallel processes.

To make the work tree function in parallel we use atomic variables alongside regular ones. For this purpose a work cell has five properties -- two atomic booleans indicating whether there has been an attempt at locking and enqueuing it, respectively, a successful lock boolean and two integers, one for thread identification and the other for the number of points in the thread's expansion set within the cell. There are four possible cell states: initial state (unclaimed), failed claim, successful claim and enqueued. The restart set is implemented as a global mutex-protected restart queue.

The expansion set of the sequential algorithm is implemented as a queue of indices of seed points within the spatial tree's point vector. In the parallel adaptation we add the index of the work cell within which this point is contained. This index is obtained during point insertion while recursively descending from the root node by checking whether the spatial tree's internal node has a valid index to the work tree, making this modification inexpensive.

We attempt to claim a cell using an adapted version of Dekker's algorithm \cite{Dijkstra1968}. First, a thread attempts to set the lock attempt flag. If it hasn't been set before, the flag is set and the thread proceeds to check all other such flags in a cell's nearest neighbours. If none of the neighbouring cells has this flag set, the successful lock flag is set, otherwise the thread attempts to set the enqueued flag and inserts the point index onto a restart queue, if it was the one to set it. In this case the state of the cell is that of a failed claim.

In a successfully locked cell, the counter of enqueued points is incremented and decremented according to how points are enqueued and dequeued from the owner thread's expansion queue. Once the number of enqueued points within a claimed cell falls to zero, the cell is unlocked by first resetting the cell's state to the initial state, and then all neighbouring cells that the thread has failed to claim are set to the enqueued state. Cells in this state may be claimed again.

Points are inserted into the spatial tree regardless of a work cell's status within which they lie. Once inserted, a check is made whether the candidate point's index is equal to the parent's and if that fails, the cell within which the candidate lies is checked for claimed status by the inserting thread. If that fails, an attempt is made to claim the cell. If the cell was successfully claimed and was previously enqueued, all points that lie within the newly claimed cell are also enqueued. This operation is made simple as both the internal spatial node and the work cell store each other's indices so full spatial tree traversal can be avoided, making its computational complexity dependent on the work tree leaf size and not the whole problem size. Lastly, if the thread was unable to claim the cell, it is enqueued onto the restart queue, if it wasn't already.

After parent point expansion, a check is made whether the front size within the parent point's cell is greater than zero. Failing that the cell is unlocked and the relevant failed claims cleared.

\begin{figure}
    \centering
    \includegraphics[width=\columnwidth]{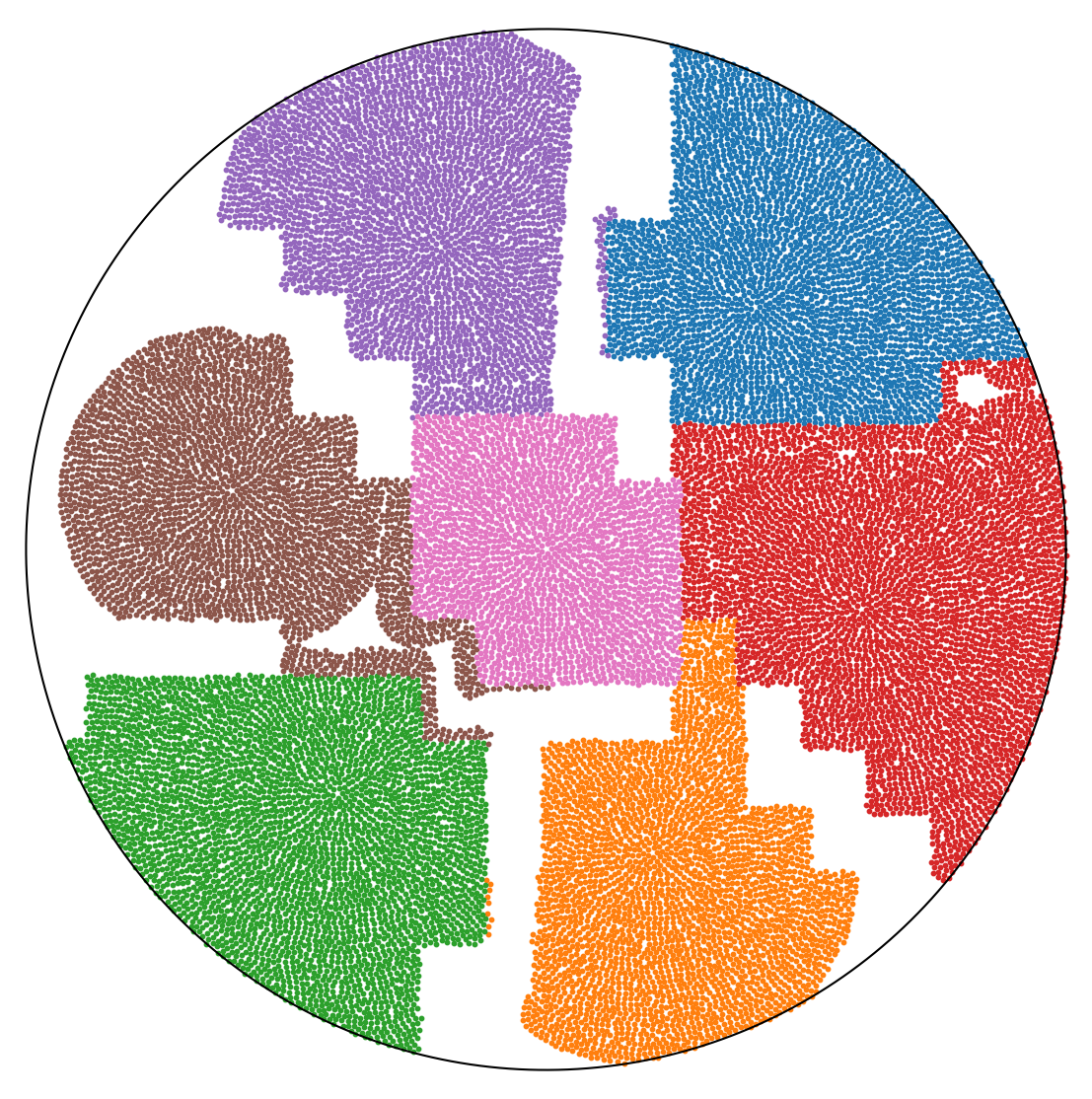}
    \caption{\label{fig:fill_demo} An example of the test domain being filled by 7 threads with $h = 0.01$ and leaf size limit set to 100 while nearing the end of the algorithm's first stage. Points of the same colour were all placed by the same thread. Threads can be seen in various stages of filling with some idling while others are in the process of backfilling seams made previously while the gaps between others are still being formed.}
\end{figure}

Intuitively, this algorithm should produce a gap of at least a single and at most two unfilled cells wide between each thread's expansion fronts after all threads exhaust their point queues (as can be seen in \cref{fig:fill_demo} between the orange and red points). This introduces the need for multiple stages of the algorithm where the fill is restarted by having threads restart the fill from work cells that have been inserted onto the restart queue, but have not been filled yet.

The statement in the previous paragraph isn't perfectly accurate as the algorithm occasionally fills the gaps after itself in a zipper-like phenomenon within a single stage and can best be seen for the thread that is placing brown points in \cref{fig:fill_demo}. This happens when one thread's front is lagging behind the other's. At some point the straggling front is able to attempt to claim cells it otherwise couldn't. These cells already contain some points which are then enqueued. On the next pass of the queue some of these points will have their candidates expand into even more cells that form the gap and so on until the whole gap (that was made until this point) is filled in. 

Nevertheless, there may be some enqueued cells left over that no thread was able to claim. For this purpose we restart the algorithm using the points from the restart queue if their respective cells are in the enqueued state. These indices are shuffled in an attempt to evenly distribute the thread's starting points within the domain. The indices are sequentially iterated over and a single cell is claimed at a time for each thread until all threads are able to restart or the indices are exhausted. The indices that did not result in a successful claim are inserted back into the restart queue to ensure that these cells are filled at some point.

\subsection{Hardware, timing and domain}

We performed all of our tests on a two-socket machine with AMD EPYC 7702 64 core processors, permitting the use of up to 256 hardware threads. The machine has \SI{512}{\giga\byte} of available memory.

Both Pfill and the proposed algorithm use \texttt{-O3} and \texttt{-march=native} compilation flags.

For timing purposes we perform only one fill at a time to prevent interference between processes, despite being able to run multiple benchmarks in parallel in terms of the number of CPU cores. We also lock threads to CPU cores unless we use more threads than there are cores.

We choose to perform our benchmark on a disc domain and set $h$ to be constant across the whole domain. We do this both to constrain the number of variables and to have a full work tree as recursive spatial tree leaf subdivision to satisfy the third condition from \cref{sec:parallelisation_concept} hasn't been implemented yet. Nevertheless, these results should be representative of the full version of the algorithm as both Pfill and our algorithm generate candidate points and perform inside checks in the same manner, thus a more complex domain with a costlier inside check (and perhaps more check failures) would lead to a similar decrease in performance in both algorithms. Introduction of variable density would lead to a differing number of cells being checked in the work cell claiming step, but otherwise our algorithm would behave very similarly in the fill phase. One such notable feature is work exhaustion from suboptimal seed point placement which is present in both algorithms and is explained in more detail in \cref{sec:staging}.

To minimise the number of unfilled cells in the first stage of the fill we choose to start each thread from a single seed point. Pfill initialises its parallel fill by performing a fill on a coarser level to generate starting seed points, but such an approach is inherently unable to guarantee an exact number of points that are evenly distributed. For this reason we choose a relaxation approach where we randomly insert points into the domain and then shift them based on their local repulsion defined by some arbitrary repulsive radial potential centred on each seed point and also originating from the edge of the domain. Like Pfill's initialisation, this additional computational load is negligible with respect to the subsequent fill and is independent of the problem complexity in terms of inserted points.

Lastly, we note that when we measure the performance of out algorithm, we don't include the tree prebuilding phase as it is currently still executed sequentially. We have measured the sequential initalisation and prebuilding times and, assuming perfect scaling, these operations should take only a few \% of the fill phase's time. Hence, the conclusions we make are nearly identical to those what would have been made with a fully parallel version of our algorithm.

\section{Results}
\label{sec:results}

Before performing scaling tests and comparing the results to Pfill we need to constrain some free parameters. The spatial tree's leaves contain a list of points whose size is a compile-time variable. We use 40 as our spatial leaf size which was recommended by \cite{Rojc2022}, but we need to determine the work cell size condition separately. 

\begin{figure}
    \centering
    \includegraphics[width=\columnwidth]{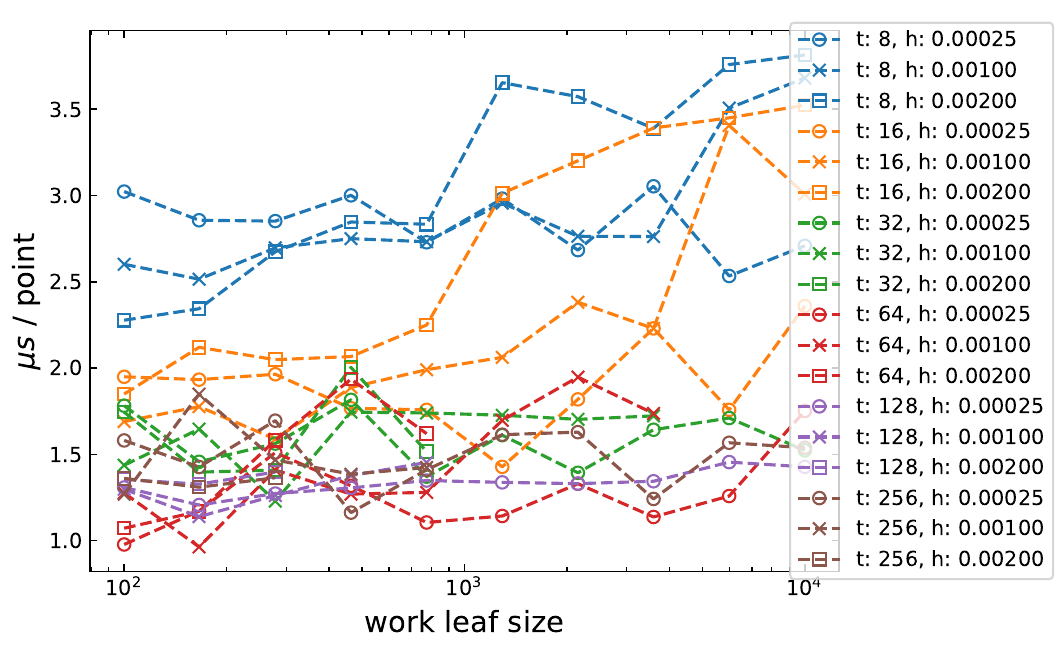}
    \caption{\label{fig:leaf_calibration} Parameter sweep for working leaf size calibration. Different colors correspond for different number of CPU cores used and the markers denote different problem sizes.}
\end{figure}

We choose not to investigate coupling of the two leaf sizes and fix the spatial one and perform a parameter scan for different problem sizes and numbers of CPU cores used. Some of these results are shown in \cref{fig:leaf_calibration} where we omit results for fewer than eight cores for legibility. Despite averaging (ranging from 5 to 10 realisations) results don't show a clear trend. Nevertheless, we can surmise that there is a general preference for very small leaf size. We emphasize that this parameter represents an upper limit on leaf size and that leaves may contain up to $1/2^d$ of points the limit is set to. 

From observing these results we set the work cell size limiting parameter to 100 for all future runs.

Next, we perform strong scaling benchmarks for both this algorithm and Pfill. We choose two constant spacing functions with $h = 0.001$ and $h = 0.00025$ where both choices produce results with a large number of points (about \num{2e6} and \num{4e7}, respectively), but the final dynamics might be different. 

\begin{figure}
    \centering
    \includegraphics[width=\columnwidth]{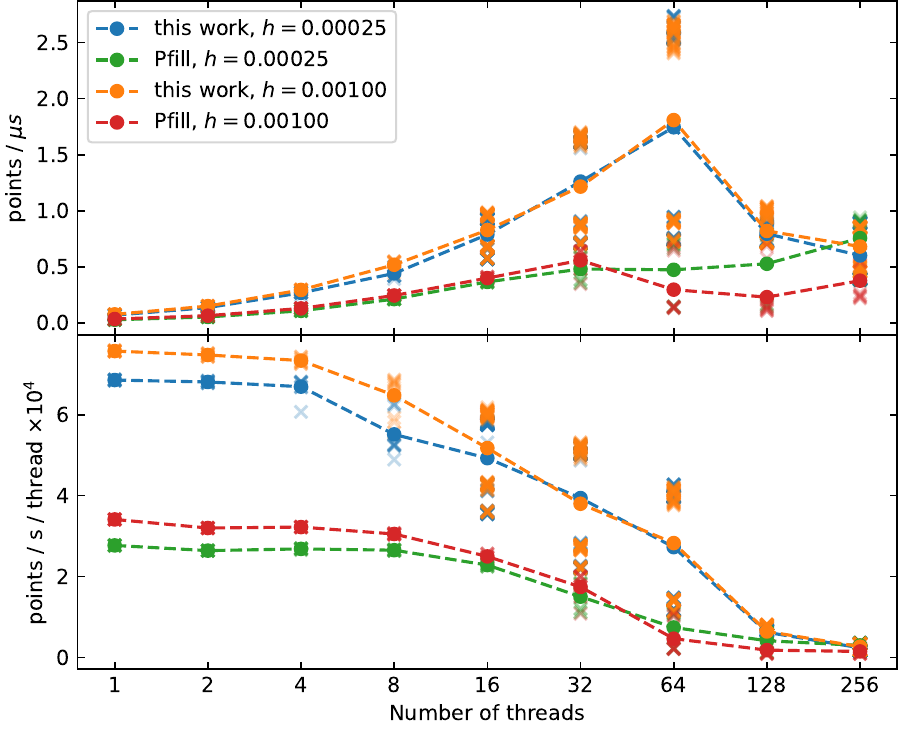}
    \caption{\label{fig:strong_scaling} Strong scaling tests for the presented algorithm and Pfill. The top panel shows the total point throughput while the bottom panel shows the throughput per thread. Markers connected by dashed lines represent averages of markers of the same color and number of threads used.}
\end{figure}

The results of this comparison are shown in \cref{fig:strong_scaling} where the top panel shows the total point insertion rate with respect to the number of threads used while the bottom panel shows this rate normalised per thread. The results in the top panel are encouraging as the markers connected by dashed lines represent performance averages that show that the presented algorithm notably outperforms Pfill in all cases but the one where all hardware threads are used. The performance of the proposed algorithm increases only up to 64 threads but then its point throughput actually decreases. The decrease up to 128 cores is most likely due to the transition from the program executing on one CPU to two on our two-socket machine. This notably increases synchronisation costs as inter-socket synchronisation of CPU cores can no longer be established through caches and suffers the same latency penalties as memory lookups. Regardless of this, a clear trend can be seen on the lower panel where the point throughput efficiency is logarithmically decreasing with an increasing number of threads leading to only $\sim\SI{20}{\percent}$ efficiency at 64 cores where the total throughput is still increasing.

We also show the samples that were used to obtain averages to expose a peculiar bifurcation of results around the average as one would otherwise intuitively expect the samples to be more concentrated near the average. These are the most pronounced at 64 threads and by observing the lower panel of \cref{fig:strong_scaling} it suggest that about a factor of two improvement in efficiency is possible.

\subsection{Staging}
\label{sec:staging}

Both Pfill and the proposed algorithm suffer from thread idling due to front exhaustion as after the thread empties its point queue, it idles until all other threads do the same. The proposed algorithm might additionally suffer from this phenomenon even more as we restart the fill multiple times, but usually the unfilled volume a relatively small fraction of the total volume, so any imbalance is dwarfed by the inefficiencies in the initial stage\footnote{Looking at the distribution of all data from \cref{fig:strong_scaling} more than half of all realisations spend at least \SI{99.4}{\percent} of the time in the first stage and for $h = 0.00025$ the lowest fraction in the first stage is \SI{99.8}{\percent}.}.

\begin{figure}
  \centering
  \includegraphics[width=\columnwidth]{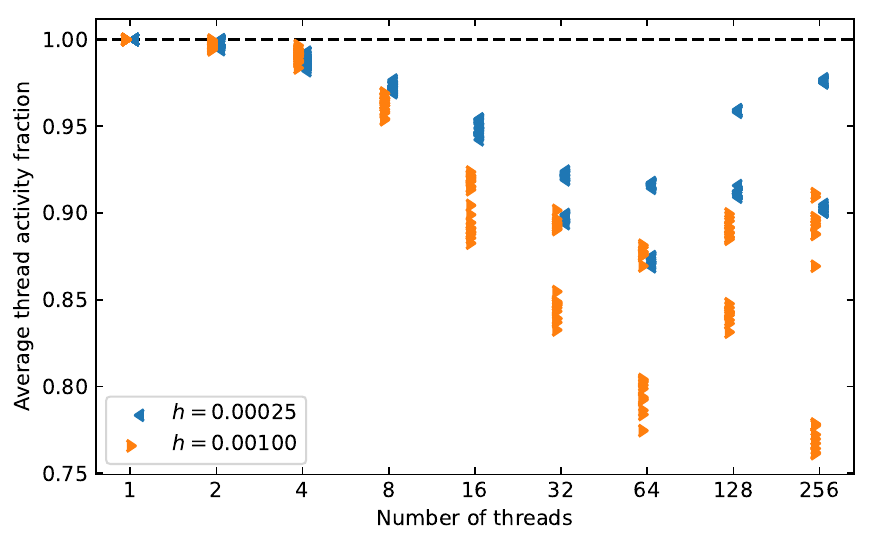}
  \caption{\label{fig:thread_activity} Average fractions of time threads spend actively computing with respect to the total computation time for different number of threads and two different problem sizes for benchmarks shown in \cref{fig:strong_scaling}. The black horizontal dashed line indicates the maximum possible thread activity while the offset carets are used for legibility when displaying data from different problem sizes at similar coordinates.}
\end{figure}

Thread idling time averages for all realisations from \cref{fig:strong_scaling} are shown in \cref{fig:thread_activity} which, show a similar bifurcation patten as in \cref{fig:strong_scaling} suggesting that this inefficiency might be at least partially responsible for it. It is unlikely it fully explains it, since the clustering in the upper panel of \cref{fig:strong_scaling} at 64 threads suggests $\sim3\times$ greater point throughput and consequently $\sim3\times$ better efficiency difference between the two clusters while the activity difference is only $\sim\SI{5}{\percent}$.

\section{Discussion}
\label{sec:discussion}

In terms of sensitivity to hyperparameters the proposed algorithm seems to be relatively insensitive to work cell size as can be seen in \cref{fig:leaf_calibration} where performance remains the same for changes in leaf size by one order of magnitude in cases with few threads and for two orders of magnitude for 32 and more threads. Here we have to note that while the condition for work tree point splitting may be continuous, the cell sizes are discrete and should actually change only when the leaf size condition changes by $2^d$. It is very likely that the optimal parameter is dimensionally dependent, but there might exist a scaling relation that provides a good initial value for any dimension once it is known for some dimension.

The multistage nature of the algorithm could, at a first glance, present a significant source of inefficiency, but it turns out that there is relatively little time spent filling cells that were left enqueued. This phenomenon might also be connected to the preference for smaller work cell sizes as greater granulation leads to fewer points needing to be inserted in subsequent stages.

Even though the bottom panel of \cref{fig:strong_scaling} makes it seem like Pfill begins to notably approach to the proposed algorithm's performance as the number of threads increases, a closer look at the data shows, that for the $h=0.00025$ case it never surpasses \SI{60}{\percent} below and including 64 threads with the biggest difference being right at this limit where the relative performance drops to \SI{35}{\percent}. We haven't been able to diagnose to what extent different factors influence the decrease in efficiency with a growing number of threads, but the way Pfill's slope changes in the bottom panel of \cref{fig:strong_scaling} to being parallel with the proposed algorithm's result, this might indicate that both encounter a similar limiting factor.

\subsection{Distributed-memory adaptation}

The work on this algorithm is done in the context of its extension to distributed-memory systems. As work cell sizes can be adjusted for specific needs, we imagine that this concept can be adapted for use in distributed-memory systems. In this setting, mutexes or atomics can no longer be used and other synchronisation mechanisms need to be implemented. One possible adaptation could use the shared-memory version of the algorithm presented in this work within each distributed process, and instead of explicitly claiming work cells these processes would communicate which coarse-level work cells they are entering to a parent process which would oversee the progression of all children. When the parent process would detect that two child processes have entered the same cell, this conflict would then be communicated to the two processes which would consolidate their work. Such an approach also allows for separate (although necessarily aligned) spatial trees that can be combined at the end of the fill process as required.

\section{Conclusion}
\label{sec:conclusion}

In this work we have presented a new parallel algorithm for filling domains with quasi-randomly distributed points with uniform density. We performed benchmarks and compared our results to the ones obtained by using a previously existing algorithm, Pfill. It is evident that the proposed algorithm provides a notable improvement in all situations where it shows favourable scaling by a factor of about two. Total throughput of points can be seen to increase with an increasing number of threads up to 64, but then it drops significantly when increased to 128, utilising both CPUs on the machine. We haven't diagnosed the source of the issues yet, but suspect that this is due to a significant increase in cache miss costs when data needs to be synchronised between CPUs instead of within them.

Another notable and yet unexplained phenomenon is the bifurcation of performance times for all benchmarks and is most pronounced for runs with a higher number of threads, being the most pronounced at 64 threads where the algorithm also produces the highest throughput of points. We have looked at average thread activity times for each benchmark realisation to asses whether work imbalance due to asymmetric front exhaustion is responsible for this, but have found that the bifurcation we observe there isn't significant enough to explain these results. From this we conclude that this asymmetry is a symptom rather than the cause of these differences. We suspect that this phenomenon originates from different memory allocation layouts between different realisations, but have yet to find a way to confirm this.

Similarly, we have also investigated the amount of time threads spend in the first an all other stages of the fill. Encouragingly, very little time is spent all but the first stage. An obvious improvement of the algorithm would be for threads to seek work when they exhaust their fronts by attempting to resume filling starting from the enqueued cells inserted into the restart queue, but judging from our results it would be more optimal for the threads to go and seek yet unfilled cells. We imagine that fill completeness information could be included in the one of the tree data structures to aid threads in seeking as distant cells as possible to resume work from. Notably, the latter approach represents a possible significant advantage over Pfill which can suffer from work imbalance between threads.

Furthermore, since both trees perform similar functions, they could be merged, leaving only the work tree leaf nodes as an additional structure of certain internal nodes of the spatial tree. Currently, the proposed algorithm's implementation is only capable of handling mild density variations, as we have yet to implement adaptive spatial leaf splitting in the tree prebuilding phase which would permit having work tree leaves at different depth levels. Lastly, we aim to investigate the causes of performance bifurcation since if we can guarantee performance of the better branch, the overall performance of the algorithm will be vastly improved.

\section*{acknowledgment}
The authors would like to acknowledge the financial support of the Slovenian Research and Innovation Agency (ARIS) research core funding No.\ P2-0095, Young Researcher programmes PR-10468 and PR-13389.
\bibliographystyle{IEEEtran}
\bibliography{references}

\begin{thebibliography}{10}
\providecommand{\url}[1]{#1}
\csname url@samestyle\endcsname
\providecommand{\newblock}{\relax}
\providecommand{\bibinfo}[2]{#2}
\providecommand{\BIBentrySTDinterwordspacing}{\spaceskip=0pt\relax}
\providecommand{\BIBentryALTinterwordstretchfactor}{4}
\providecommand{\BIBentryALTinterwordspacing}{\spaceskip=\fontdimen2\font plus
\BIBentryALTinterwordstretchfactor\fontdimen3\font minus \fontdimen4\font\relax}
\providecommand{\BIBforeignlanguage}[2]{{%
\expandafter\ifx\csname l@#1\endcsname\relax
\typeout{** WARNING: IEEEtran.bst: No hyphenation pattern has been}%
\typeout{** loaded for the language `#1'. Using the pattern for}%
\typeout{** the default language instead.}%
\else
\language=\csname l@#1\endcsname
\fi
#2}}
\providecommand{\BIBdecl}{\relax}
\BIBdecl

\bibitem{Liu2002}
G.~Liu, \emph{Mesh Free Methods: Moving Beyond the Finite Element Method}.\hskip 1em plus 0.5em minus 0.4em\relax CRC Press.

\bibitem{Oanh2017}
D.~T. Oanh, O.~Davydov, and H.~X. Phu, ``Adaptive rbf-fd method for elliptic problems with point singularities in 2d,'' vol. 313, pp. 474--497.

\bibitem{Slak2019}
J.~Slak and G.~Kosec, ``On generation of node distributions for meshless pde discretizations,'' vol.~41, no.~5, pp. A3202--A3229.

\bibitem{Fornberg2015}
B.~Fornberg and N.~Flyer, ``Fast generation of 2-d node distributions for mesh-free pde discretizations,'' vol.~69, no.~7, pp. 531--544.

\bibitem{Bacer2025}
L.~Bacer, R.~Zamolo, D.~Miotti, and E.~Nobile, ``Adaptive rbf-fd meshless solution of 3d fluid flow and heat transfer problems,'' vol. 179, p. 106367.

\bibitem{Depolli2022}
M.~Depolli, J.~Slak, and G.~Kosec, ``Parallel domain discretization algorithm for rbf-fd and other meshless numerical methods for solving pdes,'' vol. 264, p. 106773.

\bibitem{Rojc2022}
B.~Rojc and M.~Depolli, ``Parallel spatial indexing for domain discretization,'' in \emph{2022 45th Jubilee International Convention on Information, Communication and Electronic Technology (MIPRO)}.\hskip 1em plus 0.5em minus 0.4em\relax IEEE, pp. 251--256.

\bibitem{Rojc2021}
------, ``A resizable c++ container using virtual memory,'' in \emph{Proceedings of the 16th International Conference on Software Technologies}.\hskip 1em plus 0.5em minus 0.4em\relax SCITEPRESS - Science and Technology Publications, pp. 481--488.

\bibitem{Ibaroudene1991}
D.~Ibaroudene, \emph{Algorithms and parallel architecture for multidimensional image representation}.\hskip 1em plus 0.5em minus 0.4em\relax State University of New York at Buffalo.

\bibitem{Dijkstra1968}
\BIBentryALTinterwordspacing
E.~W. Dijkstra, ``Cooperating sequential processes,'' published as { EWD:EWD123pub}. [Online]. Available: \url{http://www.cs.utexas.edu/users/EWD/ewd01xx/EWD123.PDF}
\BIBentrySTDinterwordspacing

\end{thebibliography}

\end{document}